\documentclass[aps,prl,twocolumn,showpacs,superscriptaddress,floatfix]{revtex4}
\usepackage{graphicx}
\newcommand{\la}{\lambda}
\newcommand{\ep}{\epsilon}

\newcommand{\xv}{\vec{x}}
\newcommand{\be}{\begin{equation}}
\newcommand{\ee}{\end{equation}}
\newcommand{\bea}{\begin{eqnarray}}
\newcommand{\eea}{\end{eqnarray}}

\begin{document}

\title{Breakdown of Kolmogorov scaling in models of cluster aggregation
with deposition}
\author{Colm Connaughton}
\affiliation {Laboratoire de Physique Statistique de l'ENS, 24 rue Lhomond - 75231 Paris cedex 05, France}
\author{R. Rajesh}
\affiliation{
Martin Fisher School of Physics, Brandeis University,
Mailstop 057, Waltham, MA 02454-9110, USA}
\author{Oleg Zaboronski}
\affiliation{Mathematics Institute, University of Warwick, Gibbet Hill
Road, Coventry CV4 7AL, UK}

\date{\today}

\begin{abstract}
The steady state of the model of cluster aggregation with
deposition is characterized by a constant flux of
mass directed from small masses towards large masses. 
It can therefore be studied
using phenomenological theories of turbulence, such as
Kolmogorov's 1941 theory. On the other hand, the large scale
behavior of the aggregation model in dimensions lower than or equal to two
is governed by
a perturbative fixed point of the renormalization group flow, which
enables an analytic study of the scaling properties of
correlation functions in the steady state. In this paper, we
show that the correlation functions have multifractal scaling, which
violates linear Kolmogorov scaling. The analytical results are verified by
Monte Carlo simulations.
\end{abstract}


\maketitle

Understanding Navier-Stokes turbulence is a big challenge of
modern theoretical physics. It is a 
nonlinear system far from equilibrium with no obvious small
parameters that would allow a perturbative treatment.
It would
therefore be instructive to study simpler nonequilibrium models which
possess the qualitative features of turbulent systems, yet are analytically
tractable.
Such an approach has
proved to be fruitful, especially during the past
decade. The study of Burgers turbulence \cite{frischBook}, 
the Kraichnan model of
passive scalar advection and turbulence in kinematic magneto
hydrodynamics \cite{falkovich2001}
led to a
better understanding of the limitations of phenomenological
theories of turbulence. Also, it concentrated attention on
more general concepts of turbulence, such as intermittency
and breakdown of self-similarity.

In this paper we study breakdown of self-similarity 
in a model of diffusing, coagulating masses in the presence
of a steady influx of monomers.  Consider a $d$-dimensional
hypercubic lattice whose sites are occupied by particles that
carry positive masses. Multiple occupancy of a site is
allowed. Given a configuration of particles, the system evolves in
time as follows. At rate $D$, a particle hops to one of its $2d$ nearest
neighbor sites. At rate $2 \lambda$, two particles on the same site
coalesce together to form a new particle whose mass is the sum of masses of
its constituents. At rate $J/m_0$, a particle with mass $m_0$ is
injected at a site. The parameter $J$ is the average mass flux into the
system.  For simplicity of analysis it is assumed that the 
rates $D$ and $\lambda$ do not depend on the particles masses.
The initial condition is one in which there
are no masses. We call this model the mass model (MM).
A generalization of this model is one where the masses (or 'charges') could be
positive or negative. Here, with rate $J_{c}/m_0^2$ particles of charge 
$m_0$ and $-m_0$ are input into the system. $J_{c}$ has the meaning of
average influx of square of the charge. 
This model will be called the charge model (CM).
We will be interested in the continuous limit of these models.

A feature of the steady state of MM is
the presence of constant flux of mass from small masses to large masses via
coagulation. This is analogous to turbulent systems where there is a constant
flux of energy from small wave numbers to large wave numbers via nonlinear
interactions.
Let $C_{n}(m_1,\ldots,m_n) (\Delta V)^n \prod_i dm_i$ 
be the probability of having particles of masses in the intervals
$[m_i, m_i+dm_i]$ in a  volume $\Delta V$.
We ask how $C_n(m_1,\ldots,m_n)$ varies with mass when 
$m_1,\ldots,m_n \gg m_0$. In particular what is the
value of the homogeneity exponent $\gamma(n)$ defined through 
$C_n(\Gamma m_1,\ldots,\Gamma m_n)
=\Gamma^{-\gamma(n)} C_n(m_1,\ldots,m_n)$?
The quantity in turbulence that is analogous to 
$C_n(m_1,\ldots,m_n)$ would be
$\langle E_{k_1} \ldots E_{k_n} \rangle$, where $E_{k_i}$ 
is the energy corresponding to the wavenumber $k_i$.

MM has been studied in many different contexts. Examples include
submonolayer epitaxial thin film growth by deposition of atoms onto a
substrate in the limit when the distance between
clusters is much larger than typical size of a cluster \cite{KMR1999}, river
networks \cite{scheidegger1967,dodds1999}, force fluctuation in
in granular bead packs \cite{coppersmith1996} 
and nonequilibrium phase transitions \cite{MKB2000,rajesh2004}. It was
one of the first models for self organized criticality wherein power laws are
generated from simple dynamical rules. In this context, it also maps
\cite{dhar1999} onto the
abelian directed sandpile model of self organized criticality
\cite{dhar1989}. 
Finally, the mean field limit of MM is
mathematically similar to the kinetic equations of
three-wave weak turbulence \cite{zakharovBook}.

It was shown in Refs.~\cite{takayasu1989,takayasu1997,huber1991} that
in one dimension $\gamma(1)=4/3$ for MM and $\gamma(1)=5/3$ for CM. By
studying the two point correlations, it was shown that in $d<2$,
$\gamma(1) = (2 d+2)/(d+2)$ for MM and $\gamma(1) = (3d+2)/(d+2)$ for CM
\cite{majumdar1993,RM2000}.

In this paper, using the renormalization group (RG) formalism, the exponent 
$\gamma(n)$ is calculated as an expansion in $\ep=2-d$ up to order 
$\epsilon$.  In two
dimensions, the upper critical dimension of the model, the logarithmic
corrections to the mean field results for  $C_n(m_1,\ldots,m_n)$ 
are calculated. Exact results are obtained for $C_2(m_1,m_2)$ in all
dimensions. The Kolmogorov
prediction for $\gamma(n)$ based on self similarity is shown to break
down.

We first determine the dependence of  $C_n(m_1,\ldots,m_n)$ on mass 
using a self-similarity
conjecture similar to Kolmogorov's 1941 conjecture about the
statistics of velocity increments in hydrodynamic turbulence.
Assume that $C_n$ depends only on the masses $m_i$, 
mass flux $J$ and the diffusion coefficient $D$.
The dimensions of the various parameters describing the continuous limit of
the model are
$[J]=M L^{-d} T^{-1}$,
$[D]=L^{2}T^{-1}$, $[C_n]=L^{-n d} M^{-n}$ and $[m]=M$. 
There is a
unique combination of $D$, $m$ and $J$ that has the dimension of
$C_{n}$ given by
$C_{n}\sim (J D^{-1})^{n d/(d+2)} m^{-\gamma_{kolm} (n)}$,
where
\be
\gamma_{kolm}(n) = \left(\frac{2 d+2}{d+2}\right) n,
\label{eq:kolmogorov}
\ee
is the Kolmogorov scaling exponent. As expected, the dependence of 
$\gamma_{kolm}$ on the index
$n$ is linear, reflecting the assumed self-similarity of the
statistics of the local mass distribution $N(\xv,m,t)$. 
When $n=1$, $\gamma_{kolm} =(2 d+2)/(d+2)$, which agrees with the result of
an exact computation for $d<2$ \cite{RM2000}.

The self-similarity conjecture assumes that $C_n$ 
does not depend on the following: the reaction rate $\lambda$,
the lattice spacing, the position of the source $m_0$ and the box size
$\Delta V dm_1\ldots dm_n$. 
The lack of dependence on the lattice spacing
is expected due to the renormalizability
of the effective field theory describing MM below two
dimensions. We will however find an anomalous dependence of
correlation functions on a length scale depending on the other parameters
that leads to the violation of self-similarity.

Starting from the lattice model, it is possible to construct an effective
field theory of MM using the formalism due to Doi and 
Zeldovich \cite{doi1976a,zeldovich1978}. 
Furthermore, it is possible to
establish an exact map between this field theory and the following
stochastic integro-differential equation, \cite{lee1994,zaboronski2001}:
\bea && \left(\frac{\partial}{\partial t} -D \nabla^2 \right)
\phi(m) =
\la \int_{0}^{m}  dm' \phi(m') \phi(m-m')
\nonumber\\
&&\mbox{} - 2\la \phi(m) N +\frac{J}{m_{0}}\delta (m-m_{0})
+i\sqrt{2\lambda}\phi(m) \eta(\xv,t), \label{sse}
\eea
where $N=\int_{0}^{\infty} dm'\phi(m')$,
$i=\sqrt{-1}$, and $\eta(\xv,t)$ is white noise in space and time with
$\langle \eta(\xv,t) \eta(\xv',t')\rangle =\delta(t-t')\delta^d (\xv-\xv')$.
All correlation functions of the mass distribution can be
expressed in terms of the correlation functions of $\phi(m,\xv,t)$
In particular,
\bea
C_n(m)\approx \frac{1}{n!}\langle [\phi(m,\xv,t)]^n \rangle,
\eea
where $\langle \ldots \rangle$ denotes averaging with respect to noise
$\eta$  \cite{CRZunpublished}.

Without the noise term, Eq.~(\ref{sse}) reduces to the mean field
Smoluchowski equation of the model. Thus, all the fluctuation
effects are encoded in the imaginary multiplicative noise term.
Equation~(\ref{sse}) simplifies after taking Laplace transform with
respect to the mass variable \cite{zaboronski2001}. Let
$R_{\mu} (\xv,t)=N(\xv,t)-\int_{0}^{\infty} \!\!\!  dm \phi(\xv,m,t) e^{-\mu m}$.
Then,
\be
\left(\!\frac{\partial}{\partial t}\! -\! D \nabla^2\! \right)\!
R_{\mu} = -\la R_{\mu}^2+\frac{j}{m_{0}}+  i
\sqrt{2 \la}R_{\mu}\eta(\xv,t),
\label{sre}
\ee
where $j=J(1-e^{-\mu m_{0}})$, and the dependence of $R_{\mu}$ on $(\xv,t)$
has been suppressed.

Equation~(\ref{sre}) has the
form of the stochastic rate equation of the $A+A \rightarrow A$
reaction in the presence of a source \cite{droz1993}, reducing the
the computation of the mean mass distribution in MM to solving
a one-species particle problem.
However, in order to compute $C_{n}(m,t)$, 
the correlation functions of the form $\langle R_{\mu_{1}}(\xv,t)
R_{\mu_{2}}(\xv,t) \ldots R_{\mu_{n}}(\xv,t)\rangle$ need to be determined. 
These are
non-trivial, as the stochastic fields $R_{\mu}(\xv,t)$'s are
correlated for different values of $\mu$ via the common noise term
in Eq.~(\ref{sre}). 

\begin{figure}
\includegraphics[width=7.0cm]{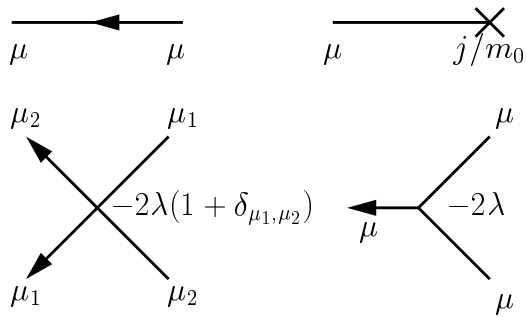}
\caption{\label{fig1} Propagators and vertices of the theory.}
\end{figure}

The 
set of Feynman rules for perturbative computation of correlation functions
$C_n$ follows from Eq.~(\ref{sre}) \cite{drouffe1994}
and is summarized in Fig.~\ref{fig1}.
The $n$-point correlation function $\langle \prod_{i=1}^n 
R_{\mu_{i}}(\xv_{i},t_{i}) \rangle$ is given by the sum of 
all Feynman diagrams that have $n$ outgoing lines 
built out of blocks shown in Fig.~\ref{fig1}.

\begin{figure}
\includegraphics[width=8.0cm]{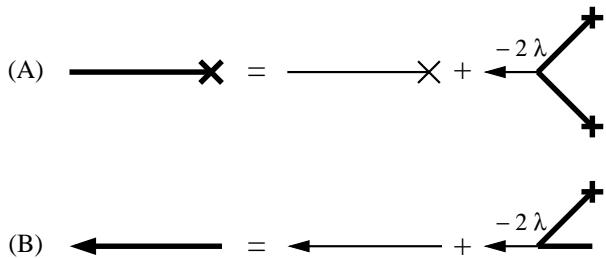}
\caption{\label{fig2} Diagrammatic form of mean field equations for: (A)
$R_{mf}$ and (B) $RR$-response function.}
\end{figure}

Let $R_{mf}$, denoted by a thick line with a cross, be the
sum of all tree diagrams with one outgoing line.
The equation satisfied by $R_{mf}$ is shown in
diagrammatic form in Fig.~\ref{fig2}A: 
$d R_{mf}/dt = j/m_0 - \lambda R_{mf}^2$. This corresponds to the noiseless
limit of Eq.~(\ref{sre}). Solving,
\be
R_{mf}(t)= \sqrt{\frac{j}{m_0 \lambda}}
\tanh\left( \sqrt{\frac{j \lambda}{m_0}} t \right)
\stackrel{t\rightarrow \infty}{\longrightarrow} \sqrt{\frac{j}{m_0
\lambda}}.
\label{rmf}
\ee

To account for the noise term, we have to include diagrams with loops.
To construct the loop expansion, it is convenient to introduce a tree level
propagator $G_{mf}(x_2 t_2;x_1 t_1)$.  The equation
obeyed by it is shown in Fig.~\ref{fig2}B. The solution is
\be
\frac
{G_{mf}({\bf 2}; {\bf 1})}
{G_{0}({\bf 2}; {\bf 1})}
= \left[\frac {\cosh \sqrt{\frac{j \lambda}{m_0}} t_1 } {\cosh
\sqrt{\frac{j \lambda}{m_0}} t_2 }
\right]^2
\stackrel{t_{1,2}\rightarrow \infty}{\longrightarrow} 
e^{-\Omega (t_2-t_1)},
\label{gmf}
\ee
where $G_{0}$ is the Green's function of the linear diffusion equation, 
${\bf 2}=(\xv_2,t_2)$,
${\bf 1}=(\xv_1,t_1)$,
and $\Omega = 2\sqrt{j\lambda/m_{0}}$ is the inverse of the mean
field response time.
All terms in the loop expansion constructed using the
vertices of Fig~\ref{fig1}, $G_{mf}$ and $R_{mf}$ are
finite in $d<2$.

We now show that the loop expansion corresponds to weak coupling 
expansion with respect to $\lambda$. 
Consider a diagram contributing to $\langle R_{\mu}^n\rangle$ that has $L$
loops, $V$ vertices and $N$ $R_{mf}$-lines. The $\lambda$ factors arise from
$L d$ momentum integrals ($\lambda^{L d/4}$), $V$ time integrals
($\lambda^{-V/2}$), $N$ $R_{mf}$ lines ($\la^{-N/2}$) and $V$ vertices
($\lambda^V$). 
Thus, the corresponding Feynman integral is proportional to
$\la^{-N/2+V/2+ L d/4}$. Also,
the number of triangular vertices in the graph equals $N-n$
and the number of quartic vertices is equal to the number of loops
$L$, i.e., $V=L+N-n$.
Thus, any $L$-loop graph contributing to the mean mass
distribution is proportional to
$\la^{-\frac{n}{2}+L (1+\frac{d+2}{4})}$. We conclude that
loop expansion corresponds to the perturbative expansion of $\langle R_\mu^n
\rangle$ around $R_{mf}^n$ with the parameter
$\la^{\frac{2+d}{4}}$.

The conditions under which the loop corrections to the tree level
answer can be neglected may be derived using dimensional
analysis. The scale of diffusive fluctuations is given by the only
constant of dimension length which can be constructed out of $D$ and
and $j/m_0$: $L_{D}= (m_0 D/j)^{1/(d+2)}$. The dimensionless expansion
parameter in the loop expansion above is therefore $g(L_D)=\la
L_{D}^{\ep}$, where $\ep=2-d$. The large mass behavior of $C_n(m)$
is determined by the small-$\mu$ behavior of $\langle R_{\mu}^n\rangle$. 
In $d<2$, $g_{0}\rightarrow \infty$ when $\mu \rightarrow 0$ and the
loop expansion breaks down. Thus, a re-summation of loop expansion
is needed to extract the $\mu\rightarrow 0$ behavior of $R_{\mu}$.
This will be done using dynamical RG formalism.

We first examine $\langle R_{\mu} \rangle$.
It was shown in Ref.~\cite{droz1993,lee1994} that 
renormalization of the coupling constant alone regularizes the loop expansion
for $\langle R_{\mu} \rangle$ when $\ep \rightarrow 0$.
In particular, $\langle R_{\mu} \rangle$ has no anomalous
dimension and therefore has the form $L_D^{-d} f[g_R,L_0/L_D]$ 
where $L_0$ is
a reference scale.  $g_R$ is the renormalized reaction
rate and  is related to $g(L_0)$ by
\be
g_R= \frac{g(L_0)}{1+g(L_0)/g^*},
\ee
where $g^*=(8 \pi)^{d/2} [2 \Gamma(\ep/2)]^{-1}$
\cite{peliti1986,lee1994}. 
$\langle R_{\mu} \rangle$ does not depend on the length scale $L_0$. Under
RG, $g_R$ goes to $g^*$ as $L_D\rightarrow \infty$
\cite{droz1993}. Hence, $\langle R_\mu \rangle \sim L_D^{-d}$.
Large $L_D$ corresponds to small $\mu$. Then $j/m_0 \approx J \mu$,
and  $R_{\mu}\sim (J \mu)^{d/(d+2)}$. The inverse
Laplace transform gives
\be
C_1(m) \sim \frac{(D^{-1} J)^{d/(d+2)}}{m^{\gamma(1)}},
\quad d<2,
\ee
where $\gamma(1)=(2 d +2)/(d+2)$. This result is exact to all orders in $\epsilon$.

The RG analysis shows that 
$C_1$ depends only on $J$, $m$ and $D$ and thus verifies the
self similarity conjecture for $n=1$. Therefore, it is not surprising that
$\gamma(1)$ coincides with $\gamma_{kolm}(1)$.

We now examine higher order correlation functions
of the form $\langle R_{\mu_1}(\xv,t) R_{\mu_2}(\xv,t) \ldots \rangle$. 
Now, extra
ultraviolet divergences, that do not get canceled by coupling constant
renormalization, appear in the perturbative expansion of these composite
operators \cite{zinnjustinBook}. These extra divergences lead to a spectrum
of anomalous dimensions which are calculated below.

\begin{figure}
\includegraphics[width=6.5cm]{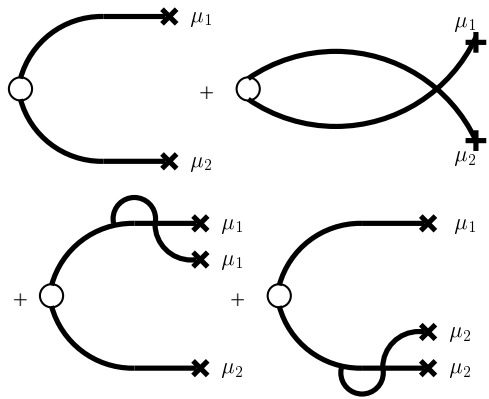}
\caption{\label{fig3} Zero and one loop Feynman diagrams contributing 
to $\langle R_{\mu_1}(\xv,t) R_{\mu_2}(\xv,t) \rangle$. 
The circles mean that all outgoing lines
terminate at the same spatial point.}
\end{figure}

The diagrams contributing to 
$\langle R{\mu_1}  R{\mu_2} \rangle$ 
up to one loop are shown in Fig.~\ref{fig3}. It is straightforward to
generalize them to $n>2$. For $n\geq 2$, diagrams contain
$n$ outgoing lines and $n\choose 2$ connected one loop diagrams.
Computing the diagrams using equations~(\ref{rmf}), (\ref{gmf})
and simplifying, we obtain
\be
\!\langle \prod_{i=1}^n \!R_{\mu_i} \rangle\! =\!\!
\prod_{i=1}^{n}\langle R_{\mu_i}\rangle \!\left[ 1
\! - \! \frac{n (n-1) \lambda \Gamma(\ep/2) } {(8 \pi D)^{d/2}
(\Omega_1\!\! +\!\! \Omega_2)^{\ep/2}}\! + \ldots \right].
\label{eq:multi}
\ee

Expressed in terms of the renormalized coupling $g_R$, Eq.~(\ref{eq:multi})
reduces to
\be
\!\langle \prod_{i=1}^n \!R_{\mu_i} \rangle\! =\!\!
\prod_{i=1}^{n}\langle R_{\mu_i}(g_R)\rangle\! \left[ 1
\! -\frac{n(n-1)g_R}{4 \pi \ep}
\!+\! O(g_R^2)\right].
\ee
Since coupling constant renormalization removes all singularities in $\langle
R_\mu(g_R) \rangle$, there are no singularities in the product
$\prod_{i=1}^{n}\langle R_{\mu_i}(g_R)\rangle$ when $\ep\rightarrow 0$. 

The remaining singularity is canceled by multiplicative renormalization of
the composite operators.  Let $
\langle \prod_{i=1}^n R_{\mu_i} \rangle_R = Z_n  \langle \prod_{i=1}^n 
R_{\mu_i} \rangle$,
where $Z_n$ is fixed by the condition that $\langle \prod_{i=1}^n R_{\mu_i}
\rangle_R$ is not singular in $\ep$ \footnote{The problem of renormalization
of composite operators is simple for MM because 
operators of lower order do not contribute to the 
renormalization of operators of higher order. The absence of mixing
is due to the absence of vertices increasing the number of 
outgoing lines in any Feynman diagram}.
Thus
$Z_n = 1+n (n-1) g_R/(4 \pi\ep) + O(\ep^2)$.
The bare $\langle \prod_{i=1}^n R_{\mu_i} \rangle$ is independent
of the reference scale $L_0$. Thus,
\be
L_0 \frac{\partial}{\partial L_0} \left[Z_n^{-1} \langle \prod_{i=1}^n 
R_{\mu_i} \rangle_R\right] =0.
\label{eq:CS}
\ee
Let $L_i=[m_0 D /j(\mu_i)]^{1/(d+2)}$. By
dimensional analysis, $\langle \prod_{i=1}^n R_{\mu_i} 
\rangle = L_0^{-2 d} f(L_i/L_0,g_R)$,
where $f$ is a dimensionless scaling function.
Substituting into Eq.~(\ref{eq:CS}), and solving the resulting equation in
the limit $L_i\gg L_0$, we obtain
$\langle \prod_{i=1}^n R_{\mu_i} \rangle_R \sim \Phi_1 
\prod_{i=1}^n L_i^{-d-\ep(n-1)/2} $, where $\Phi_1$ is a scaling function of
the variables $L_i/L_j$.
An inverse Laplace transform gives
\be 
C_n(m_1,\ldots,m_n) \sim  \Phi_2
\! \prod_{i=1}^n 
\frac{1}{m_i}\! \left(\frac{J}{D m_i}\right)^{\frac{d+\ep(n-1)/2}{d+2}}
\label{eq:cn}
\ee
$\Phi_2$ is a scaling function of the variables
$m_i/m_j$ and the parameters $\Delta V$, $m_0$ and
$\lambda$. In the limit when $m_0\rightarrow 0$ and $\lambda\rightarrow
\infty$, we expect the $\Delta V$ dependence of $\Phi_2$ to be
$\Phi_2\sim (\Delta V)^{\frac{\ep n (n-1)}{2 d}}$. 
We then conclude that
\be
\gamma(n) = \left(\frac{2 d+2}{d+2} \right)n + \left(\frac{\ep}{ d + 2}
\right) \frac{n (n-1)}{2} + O(\ep^2).
\label{eq:gamma}
\ee

The first term coincides with $\gamma_{kolm}$ [see Eq.~(\ref{eq:kolmogorov})].
The presence of the second term leads to breakdown of self similarity. This
is due to the effective anticorrelation between the particles. The self
similarity conjecture for MM 
is equivalent to a renormalized mean field theory in which only the coupling
renormalization is taken into account. Therefore, it cannot take into
account correlations between particles.

Though, Eq.~(\ref{eq:gamma}) is an $\ep$-expansion,
it is still possible to confirm that $\gamma (n)\neq \gamma_{kolm}(n)$
in $d<2$ by computing $\gamma(2)$ exactly. From the definition of
$\gamma(2)$, it follows that
$\langle
R_{\mu_{1}} R_{\mu_{2}} \rangle=(\mu_{1}\mu_{2})^{\gamma(2)/2-1}
\psi( \mu_{1}/\mu_{2})$,
where $\psi(x)$ is an unknown scaling function with the property $\psi(x) =
\psi(1/x)$.
We need to know the $\mu_{1}, \mu_{2} \rightarrow 0$ behavior of $\langle
R_{\mu_{1}} R_{\mu_{2}} \rangle$. Averaging Eq.~(\ref{sre})
with respect to noise and setting $\partial_{t} \langle R_{\mu} \rangle=0$ 
in the large time limit, we find that $\langle
R_{\mu} R_{\mu} \rangle =j/(\lambda m_{0}) \approx
J\mu/\lambda$ for $\mu \ll m_{0}^{-1}$. Comparing this result with
the above scaling form,  we find that $\gamma (2)=3$ exactly, 
which coincides with
Eq.~(\ref{eq:gamma}) with terms of $O(\epsilon^2)$ and higher order
set to zero.

Note that $\gamma(2)$ does not depend on the dimension $d$.
The result $\gamma(2)=3$ is a counterpart of the $4/5$-th law of
Navier-Stokes turbulence and is due to conservation of mass flux. 
Recall that the $4/5$ law states
that, in the inertial range the third order structure point function of
velocity field scales as the first power of
separation. While the Kolmogorov
theory respects $4/5$-law in Navier-Stokes turbulence, it
violates $\gamma(2)=3$ law in MM.

\begin{figure}
\includegraphics[width=7.5cm]{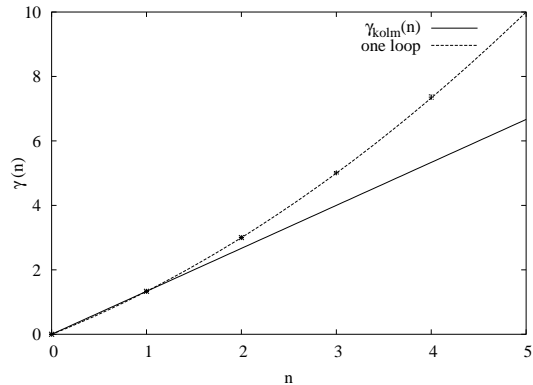}
\caption{\label{fig5} The variation of homogeneity exponent $\gamma(n)$ 
with $n$ is shown for one dimension. The dotted line corresponds to
Eq.~(\ref{eq:gamma}) with $\ep=1$ and terms of order $\ep^2$ and 
higher set to zero.
The values of $\gamma(0)$, $\gamma(1)$ and $\gamma(2)$ are exact while
$\gamma(3)$ and $\gamma(4)$ were obtained by Monte Carlo simulations.
The simulations were performed on a lattice of size $10^5$
and averaged $2\times 10^7$ times with $J=4 D$.}
\end{figure}

The terms $O(\ep^2)$ and higher in Eq.~(\ref{eq:gamma}) are either very small
or equal to zero when $d=1$. We showed their absence for $n=1,~2$ by
means of an exact computation. For $n=3,~4$ we verify their smallness
in one dimensions using Monte Carlo simulations (see   
Fig.~\ref{fig5}).

In two dimensions, logarithmic corrections to the Kolmogorov 
scaling are expected. These can be calculated
exactly using the RG method. 
We present the final results here:
\be
C_n(m)\! \sim \!\frac{J^{n/2} [\ln(m)]^{n-n^2/2}}{m^{3 n/2}}
\!\left[\!1\!+\!O\!\left(\!\frac{1}{\ln(m)} \!\right)\! \right]\!,~d\!=\!2.
\ee
The logarithmic corrections vanish for $n=2$. This is
consistent with the exact result for $C_2(m)\sim m^{-3}$, which is valid in all
dimensions.

It is straightforward to extend the above results to CM. Kolmogorov theory
for the charge model is developed using the flux of charge squared $J_c$ as 
a self-similarity parameter. This leads to Kolmogorov spectrum for 
CM, $\gamma_{kolm}^{CM} =\left(\frac{3 d+2}{d+2} \right)n$. The
stochastic equation
for CM is the same as Eq.~(\ref{sre}) modulo a redefinition of the source term:
$j/m_{0} \rightarrow 2J_{c}(1-\cosh(\mu m_{0}))/m_{0}^2)$. Therefore the 
RG analysis presented for MM
goes through with just one modification: the scale of diffusive fluctuations
in the charge model is $L_{D}=(J_{c}D^{-1}\mu^{-2})^{\frac{1}{d+2}}$ 
for $\mu\rightarrow 0$.  The result of the analysis is
\be
\gamma^{CM}(n) = \gamma^{CM}_{kolm}(n) + \left(\frac{\ep}{d + 2}
\right) n (n-1) + O(\ep^2).
\label{eq:gammacm}
\ee
In complete analogy with MM case, $O(\ep^2)$ terms in the above answer 
vanish for $n=1$ and $n=2$. Kolmogorov theory correctly predicts the 
scaling of average charge
distribution in CM, but fails to predict the scaling of multi point 
probability distributions of charge. Relation $\gamma^{CM}(2)=4$ reflects
conservation of flux of charge squared and is satisfied in all dimensions.

To summarize, we considered two models of aggregation with input which had
features qualitatively similar to turbulence. For these models, we were able to
compute the statistics of the local mass (charge) distribution and 
compare the results
against the predictions from Kolmogorov theory. It turned out that
Kolmogorov self-similarity conjecture is equivalent to a 
renormalized mean field theory approximation which
correctly accounts for effects of reaction rate
renormalization in both MM and CM. As a result, the answers for average
mass (charge) distribution derived from Kolmogorov theory turned out 
to be correct.
However, Kolmogorov theory failed to predict multi point distributions of 
mass (charge) correctly: small scale correlations
between diffusing, coagulating particles are relevant for the 
multi point statistics, 
but cannot be accounted for by coupling constant
renormalization alone.


We would like to thank G. Falkovich for useful discussions.
RR acknowledges support from NSF grant DMR-0207106. 
CC acknowledges support from Marie Curie grant HPMF-CT-2002-02004. 
OZ thanks the hospitality at Brandeis University, where
part of this work was done.

\end{document}